\documentclass[multphys,vecphys]{svmult}

\usepackage{makeidx}     
\usepackage{graphicx}    
\usepackage{multicol}    

\makeindex             

\begin{document}
\newcommand\farcs{\mbox{$.\!\!^{\prime\prime}$}}%

\title*{Optical, Ultraviolet, and Infrared Observations of SN~1993J}
\author{Alexei V. Filippenko\inst{1}\and
Thomas Matheson\inst{2}}
\institute{Department of Astronomy, University of California, 
601 Campbell Hall, Berkeley, CA  94720-3411, USA
\texttt{alex@astron.berkeley.edu}
\and Harvard-Smithsonian Center for Astrophysics, 60 Garden Street,
Cambridge, MA  02138, USA \texttt{tmatheson@cfa.harvard.edu}}
%
%
\maketitle

\abstract

We review the existing set of optical/UV/IR observations of Supernova 1993J,
concentrating heavily on optical data because these are by far the most
plentiful. Some results from theoretical modeling of the observations are also
discussed. SN 1993J has provided the best observational evidence for the
transformation of a SN from one spectral type to another, thereby providing a
link between Type II and Type Ib supernovae (SNe).  This has strengthened the
argument that SNe~Ib (and, by extension, SNe~Ic) are core-collapse events. SN
1993J has remained relatively bright for 10 years; its late-time emission comes
from the collision of supernova ejecta with circumstellar gas that was released
by the progenitor prior to the explosion.  The circumstellar material shows
strong evidence of CNO processing.

\section{Introduction}

Supernova (SN) 1993J was visually discovered in the nearby galaxy M81 (NGC
3031; $d$ = 3.6 Mpc; Freedman et al. 1994) by Francisco Garcia on 1993 March
28.906 UT (Ripero, Garcia, \& Rodriguez 1993). Ten years later, we
still vividly remember our excitement after receiving by fax a rough
finding chart from the discoverers, via the AAVSO. Ever since SN
1987A, those of us in the northern hemisphere had been eagerly waiting
for a supernova in M31, or (more likely) in a somewhat more distant
galaxy.  What a golden opportunity!  Unfortunately, our own robotic SN
search, at that time being conducted with a 0.76-m telescope at
Leuschner Observatory, missed discovery of SN 1993J because
its field of view was somewhat too small to include the position of
the supernova when centered on the galaxy nucleus, and because of bad
weather (Filippenko 1993). It was a pleasure for both of us to finally
meet Mr. Garcia, at this conference.

SN 1993J reached $V = 10.8$ mag (e.g.,
Richmond et al. 1994), becoming the brightest SN in the
northern hemisphere since SN 1954A ($m_{\rm{pg}} = 9.95$; Wild 1960;
Barbon, Ciatti, \& Rosino 1973).  In terms of observational coverage,
both in temporal consistency (almost nightly observations at early
times) and in the details of individual observations (including
observations with signal-to-noise ratios, spectral resolutions,
and wavelength regions not typically found in studies of supernovae), SN
1993J is surpassed only by SN 1987A.  Early spectra showed an almost
featureless blue continuum, possibly with broad, but weak, H$\alpha$
and He~I $\lambda$5876 lines.  This led to a Type II classification
(Filippenko et al. 1993; Garnavich \& Ann 1993; see
Filippenko 1997 for a general discussion of SN types).  Wheeler \&
Filippenko (1996) present a thorough review of the early work on SN
1993J.

Both the spectra and the light curves of SN 1993J quickly began to
indicate that this was not a typical Type II SN.  Indeed, the
initially unusual light curves and the appearance of He~I lines in the
spectra were interpreted as evidence that SN 1993J was similar to a 
SN~Ib, with a low-mass outer layer of hydrogen that gave the early
impression of a SN II (see discussion and references below).
Following Woosley et al. (1987), it was described as a ``Type IIb''
SN.  This transformation from SN~II to nearly SN~Ib indicates a common
mechanism (core collapse) for these two observationally defined
subclasses.  SN 1993J is thus one of the most significant SNe ever
studied, not only for its role in linking Types II and Ib (and
possibly Ic), but also because it was observed with such great detail.

\begin{figure}
\hspace{0.3in}
\rotatebox{270}{
\centering
\includegraphics[height=10.0cm]{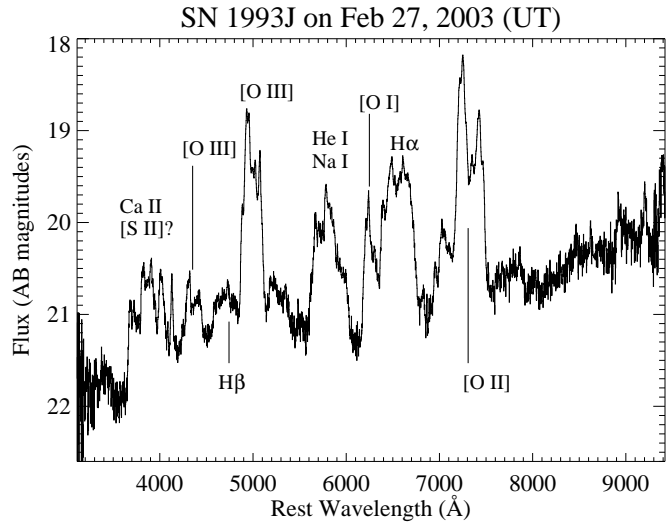}
}
%
%
\caption{Keck-I LRIS (Oke et al. 1995) spectrum of SN 1993J,
obtained by A. V. Filippenko and R. Chornock nearly 10 years after the
outburst. The absolute flux scale is only approximate, but relative fluxes
should be accurate. Plausible line identifications are labeled. AB mag = $-2.5
\, {\rm log}\, f_\nu - 48.6$, on the ordinate scale.}
\label{fig:1}       
\end{figure}

Here we review observations of SN 1993J at optical, ultraviolet (UV),
and infrared (IR) wavelengths, concentrating on optical because of the
vast amount of data obtained in this wavelength range. Because of time
and space constraints, we exclude light echoes and interstellar
absorption lines, which have been used to probe the interstellar
medium near SN 1993J.  A recent optical spectrum of SN 1993J, obtained
with the Keck-I 10~m telescope 9 years and 11 months after the
explosion, is shown in Figure \ref{fig:1}. This is very close to 10
years, so we shall dub this the official ``Valencia spectrum of SN
1993J.''

\section{Photometry of SN 1993J}

The evolution of the light curve of SN 1993J did not follow either of
the two typical paths for SNe~II.  SN 1993J did not remain at a
relatively constant brightness after a slight decline from maximum, as
a normal Type II plateau SN would, nor did the brightness decline in
the pattern of a Type II linear SN. (For representatives of these Type
II light curves, see, for example, Doggett \& Branch 1985.) Instead,
SN 1993J rose quickly, then rapidly declined for $\sim$1 week, only to
brighten a second time over the next two weeks.  This led to another
rapid decrease in brightness for $\sim$3 weeks, followed by an
approximately exponential decline.  For a complete discussion of the
photometry of SN 1993J, see Okyudo et al. (1993), Schmidt et
al. (1993), van Driel et al. (1993), Wheeler et al. (1993), Benson et
al. (1994), Lewis et al. (1994), Richmond et al. (1994, 1996), Barbon
et al. (1995), Doroshenko et al. (1995), and Prabhu et al. (1995).

The unusual initial behavior of the light curve rapidly led many SN
modelers to conclude that SN 1993J was the result of a core-collapse
explosion in a progenitor that had lost a significant fraction of its
hydrogen envelope, leaving only $\sim$0.1--0.5 $M_{\odot}$ of hydrogen.
The original envelope could have been lost through winds (H\"oflich,
Langer, \& Duschinger 1993) from a fairly massive star (25--30
$M_{\odot}$).  Another possibility explored by Hashimoto, Iwamoto, \&
Nomoto (1993; see also Nomoto et al. 1993) is that SN 1993J was the
result of the explosion of an asymptotic giant branch star having
main-sequence mass $M_{ms} \approx 7$--10 $M_{\odot}$, with a helium-rich
envelope.

A more likely solution is that the progenitor of SN 1993J was a member
of a binary system and the companion had stripped away a considerable
amount of hydrogen.  The progenitor was observed during prior studies
of M81.  Aldering, Humphreys, \& Richmond (1994) analyzed several sets
of pre-existing images and deduced that the photometry was
inconsistent with a single star at the position of SN 1993J.  They
found that the best fit for the progenitor itself was a K0~I star with
$M_{\rm{bol}} \approx -7.8$ mag and $V-R \approx 0.7$ mag.  Cohen,
Darling, \& Porter (1995) derived a similar color from a five-month
series of images of M81 from 1984; there was no apparent variability.

Using the scenario of a star that had been stripped of most of its
hydrogen envelope, Nomoto et al. (1993) and Shigeyama et al. (1994)
found a best fit to the light curve from their model of a $4 M_{\odot}$
helium core, although a range of 3--6 $M_{\odot}$ for the core is
reasonable.  The main-sequence mass of the star would have been 15
$M_{\odot}$, while the residual hydrogen envelope is less than $\sim
0.9 M_{\odot}$.  Starting with a star of initial mass of 13--16
$M_{\odot}$, Woosley et al. (1994) could reproduce the light curve from
the explosion of a remaining helium core with mass $4.0 \pm 0.5
M_{\odot}$ and hydrogen envelope with mass $0.20 \pm 0.05 M_{\odot}$.  A
similar model by Podsiadlowski et al. (1993) had $0.2 M_{\odot}$ of
hydrogen remaining on a star of initial mass $M_i \approx 15
M_{\odot}$.  Ray, Singh, \& Sutaria (1993) also invoked a binary system
for SN 1993J with a residual hydrogen envelope mass of $0.2 M_{\odot}$.
Utrobin (1994) used an envelope mass of $0.1 M_{\odot}$ remaining on a
$3 M_{\odot}$ helium core from an initial mass of $12 M_{\odot}$.
Bartunov et al. (1994) achieved a good fit to the light curve with a
helium core mass of 3.5 $M_{\odot}$, but a larger hydrogen envelope
($M_{env} \approx 0.9 M_{\odot}$).  Later studies continued to conclude
that a low-mass envelope of hydrogen on a helium core was the most
likely scenario for the progenitor (Young, Baron, \& Branch 1995;
Utrobin 1996).  Intercomparison of two methods also indicated that the
results were robust (Blinnikov et al. 1998).  Houck \& Fransson (1996)
used a non-local thermodynamic equilibrium (NLTE) synthetic spectrum
code to fit nebular spectra and found that the Nomoto et al. (1993)
models could explain the late-time spectra.  They found a best fit
with a 3.2 $M_{\odot}$ helium core with a 0.2--0.4 $M_{\odot}$ hydrogen
envelope.  Patat, Chugai, \& Mazzali (1995) also used the late-time
spectra, specifically the H$\alpha$ line, to derive an ionized
hydrogen mass of 0.05--0.2 $M_{\odot}$; this is a lower limit to the
envelope mass.

\section{Spectroscopy of SN 1993J}

Woosley et al. (1987) had already considered the above possibility for
core-collapse SNe, giving them a new name: SNe IIb.  The low-mass outer layer
of hydrogen would give the initial appearance of a SN II, but the spectrum
would slowly change to one more similar to that of a SN Ib, dominated by helium
lines with the hydrogen either appearing weakly or completely gone.  Indeed,
Nomoto et al. (1993) predicted that the spectrum of SN 1993J would show this
behavior.  This was first confirmed by Filippenko \& Matheson (1993), followed
rapidly by Schmidt et al. (1993) and Swartz et al. (1993).  Studies of the
early optical spectra include those of Filippenko, Matheson, \& Ho (1993),
Wheeler et al. (1993), Taniguchi et al. (1993), Garnavich \& Ann (1994), Ohta
et al. (1994), Prabhu et al. (1995), and Metlova et al. (1995).

Jeffery et al. (1994) present an early UV spectrum of SN 1993J taken
with the {\it Hubble Space Telescope (HST)} on 1993 April 15 UT.  The
other core-collapse SNe that had been observed in the UV to that point
were compared with SN 1993J, and there were striking differences.  SN
1993J had a relatively smooth UV spectrum and was more similar to SN
1979C and SN 1980K, both of which are radio sources and thus likely to
have thick circumstellar envelopes (e.g., Weiler et al. 1986).  The UV
spectra of SN 1987A, in contrast, showed broad absorption features.
The illumination from circumstellar interaction may reduce the
relative strengths of line features compared to the continuum and thus
produce the featureless UV spectra of SNe 1979C, 1980K, and 1993J
(Branch et al. 2000).

SN 1993J then evolved fairly rapidly into the nebular phase.  The
nebular-phase spectra were similar to those of a typical SN~Ib, but
the hydrogen lines never faded completely.  In fact, H$\alpha$ began
to dominate the spectrum at late times, certainly the result of
circumstellar interaction.  There were several other papers that
considered the nebular-phase spectra (and some relatively late-time
spectra).  Filippenko, Matheson, \& Barth (1994) show the transition
to the nebular phase with spectra from Lick Observatory.  Lewis et
al. (1994) present the complete La Palma archive covering days 2
through 125.  Li et al. (1994) discuss the nebular-phase spectra
observed from the Beijing Astronomical Observatory.  Barbon et
al. (1995) show the first year of observations from Asiago; the
transformation of the SN from Type II to IIb is evident, as is the
return of H$\alpha$ at late times (by $\sim$ 200 days).  A longer
baseline ($\sim 500$ days) for the spectra is found in the work of
Finn et al. (1995).

There were optical spectropolarimetric observations of SN 1993J.
Trammell, Hines, \& Wheeler (1993) found a continuum polarization of
$P = 1.6\% \pm 0.1\%$ on day 24 (assuming 1993 March 27.5 UT as the
explosion date).  Trammell et al. (1993), as well as later
considerations of the same data (H\"oflich 1995; H\"oflich et
al. 1996), argued that this polarization implied an overall asymmetry,
but the source of this asymmetry was undetermined.  The presence of SN
1993J in a binary system was implicated as a potential source for the
asymmetry.  With more epochs of observation, Tran et al. (1997) also
found a polarization in the continuum of $\sim$ 1\%, but a different
level for the interstellar polarization.  Nevertheless, they also
concluded that SN 1993J was asymmetric.  It is interesting to note
that a subsequent SN IIb, SN 1996cb, showed substantially similar
polarization of its spectra (Wang et al. 2000).

The analysis of individual aspects of the spectra has yielded some
interesting results.  Both Wang \& Hu (1994) and Spyromilio (1994)
found evidence for clumpy ejecta with blueshifted emission lines.
Houck \& Fransson (1996) argue that the lines are not actually
blueshifted, but that contamination from other lines appears to shift
them.  Nonetheless, the lines do show substructure that indicates
clumpy ejecta.

Models of the early spectra could reproduce their overall spectral shape, but
the line strengths were problematic.  Baron et al. (1993) found a photospheric
temperature of $\sim$8000 K for day 10, but the predicted hydrogen and helium
lines were too weak, possibly indicating unusual abundances or non-thermal
effects.  A later analysis including the {\it HST} UV spectrum was fit well by
including enhanced helium abundance and NLTE effects (Baron, Hauschildt, \&
Branch 1994).  Jeffery et al. (1994) also had difficulties fitting line
strengths for transitions that are susceptible to NLTE effects.  Clocchiatti et
al. (1995) studied the early spectra to follow the evolution of color
temperature and to calculate a distance to M81 ($\sim$ 3.5 Mpc) using the
expanding photosphere method (e.g., Eastman, Schmidt, \& Kirshner 1996).  The
NLTE treatment of calcium is explored by Zhang \& Wang (1996), who found a best
fit with a reduced calcium abundance.

\section{Recent Studies of SN 1993J}

Most of the above papers were published before year 2000, but in this section
we discuss several more recent studies.

Matheson et al. (2000a) present a series of 42 Lick and Keck low-resolution
optical spectra of SN 1993J from day 3 after explosion to day 2454, as well as
one Keck high-dispersion spectrum from day 383. The spectra are studied in
detail by Matheson et al. (2000b).  Spectra during the nebular phase, but
within the first two years after explosion, exhibit small-scale structure in
the emission lines of some species, notably oxygen and magnesium, showing that
the ejecta of SN 1993J are clumpy.  On the other hand, a lack of structure in
emission lines of calcium implies that the source of calcium emission is
uniformly distributed throughout the ejecta.  These results are interpreted as
evidence that oxygen emission originates in clumpy, newly synthesized material,
while calcium emission arises from material pre-existing in the atmosphere of
the progenitor (Li \& McCray 1992, 1993).  Spectra spanning the range 433--2454
days after the explosion show box-like profiles for the emission lines, clearly
indicating circumstellar interaction in a roughly spherical shell.  This is
interpreted within the Chevalier \& Fransson (1994) model for SNe interacting
with mass lost during prior stellar winds.  At very late times, the emission
lines have a two-horned profile, implying the formation of a somewhat flattened
or disk-like structure that is a significant source of emission.

Matthews et al. (2002) conducted IR photometry (windows in the interval
1.25--3.7 $\mu$m) and IR spectroscopy (windows in the interval 1.2--2.4
$\mu$m) of SN 1993J at early times, through about day 250.
As in the case of the optical bands, the IR brightness rose to a
secondary maximum and then dropped exponentially. However, the $L^\prime$ (3.7
$\mu$m) light curve exhibited an excess, beginning at day 130, which
Matthews et al. interpret as thermal emission from dust, as in the SNe~II 1987A
and 1998S. At early times, during the rise to the secondary maximum, the
spectral energy distribution (SED) of SN 1993J could be fitted with black
bodies, but such fits were too broad for the observed SEDs during the
exponential decline. Though initially featureless, the IR spectra
subsequently (during the exponential decline) became dominated by H, He, and
probably Fe line emission.

The {\it HST} SINS collaboration (Supernova INtensive Study) obtained a series
of UV spectra of SN 1993J over the course of about 7 years (C. Fransson et
al. 2004, in preparation).  It is quite clear that at late times, the emission
is almost entirely coming from the shock interaction between the ejecta and
circumstellar gas. The model fit to the observed spectrum is quite good, and
its total luminosity is fixed by the observed X-ray emission from the shock --
it is not a free parameter.  Moreover, the detailed line profiles show that the
emission is coming from a finite shell rather than a centrally peaked
distribution of gas.  There are strong, broad emission lines of N~II], N~III],
and N~IV], but only weak lines of C~III] and C~IV --- characteristic of CNO
processing. Indeed, a model spectrum having C:N = 1:13 by number in the
circumstellar gas (with which the ejecta collide) gives a relatively good fit
to the observations. CNO abundance ratios have previously been measured for a
number of SNe~II, with similar (but perhaps less extreme) results.

>From {\it HST} images with $0{\farcs}05$ resolution, Van Dyk et al. (2002)
identify four stars brighter than $V = 25$ mag within $2{\farcs}5$ of SN 1993J
that contaminated previous ground-based brightness estimates for the supernova
progenitor.  Correcting for the contamination, they find that the energy
distribution of the progenitor is consistent with that of an early K-type
supergiant star with $M_V \approx -7.0 \pm 0.4$ mag and an initial mass of
13--22 $M_\odot$.  The brightnesses of the nearby stars are sufficient to
account for the excess blue light seen from the ground in pre-explosion
observations (Aldering et al. 1994).  Therefore, the SN 1993J progenitor did
not necessarily have a blue companion, although by 2001, fainter blue stars are
seen in close proximity to the supernova.  These observations do not strongly
limit the mass of a hypothetical companion.  A blue dwarf star with a mass up
to 30 $M_\odot$ could have been orbiting the progenitor without being detected
in the ground-based images.

Explosion models and observations show that the SN 1993J progenitor had a
helium-rich envelope.  To test whether the helium abundance could influence the
energy distribution of the progenitor, Van Dyk et al. (2002) calculated model
supergiant atmospheres with a range of plausible helium abundances.  The models
show that the pre-supernova colors are not strongly affected by the helium
abundance longward of 4000~\AA, and abundances ranging between solar and 90\%\
helium (by number) are all consistent with the observations.

Recent optical spectra of SN 1993J (e.g., Figure \ref{fig:1}) show
that the optical continuum is still very bright ($B \approx V \approx
21$ mag), certainly from the circumstellar interaction. Van Dyk et
al. (2002) suggest that the putative, blue companion star has $B > 23$
mag and $U > 22$ mag, so perhaps the most likely
spectral region in which it might be observed is in the $U$ band,
where the actual observed continuum has $U \approx 22$ mag (Figure
\ref{fig:1}). At this conference, Stephen Smartt and collaborators
reported the possible detection of hydrogen Balmer lines from the
putative companion. Given the complexity of the SN spectrum (blends of emission
lines, uncertain continuum level), however, this must be verified with future observations;
most or all of the claimed Balmer lines may have other
explanations. It is also not clear whether the Balmer lines, if real,
come from a physical companion star or from an unrelated star along
the same line of sight; the {\it HST} images shown by Van Dyk et
al. (2002) reveal several possible contaminants.

\section{Final Remarks}

Although SN 1993J has provided the best observational evidence for the
transformation of a SN from one type to another, there have been other
examples.  The early spectra of SN 1987K showed hydrogen lines, but
the late-time spectra more closely resembled those of SNe Ib
(Filippenko 1988).  The transition itself was not observed, occurring
while SN 1987K was in conjunction with the Sun.  SN 1996cb underwent a
very similar metamorphosis from SN II to SN Ib; Qiu et al. (1999)
present a complete spectroscopic record of the transformation. A number of
other examples of genuine SNe~IIb have been found in recent years. In
addition, there were some suggestions of hydrogen in spectra of the
Type Ic SN 1987M (Jeffery et al. 1991; Filippenko 1992) and the SN Ic
1991A (and perhaps SN Ic 1990aa; Filippenko 1992).  SN 1993J is
clearly a significant object in the study of SNe.  By providing a link
between SNe~II and SNe~Ib, it has strengthened the argument that 
SNe~Ib (and, by extension, SNe~Ic) are also core-collapse events.

\section{Acknowledgements}

A.V.F.'s research on SN 1993J is currently supported by NSF grant AST-0307894,
as well as by NASA grants AR-9529 and GO-9114 from the Space Telescope Science
Institute, which is operated by AURA, Inc., under NASA contract NAS5-26555. We
thank the conference organizers for partial travel funds. We are grateful to
the staffs of the Lick and Keck Observatories for help with the observations.
The W. M. Keck Observatory is operated as a scientific partnership among
Caltech, the University of California, and NASA; the Observatory was made
possible by the generous financial support of the W. M. Keck Foundation.

\printindex
\end{document}